

\font\mybb=msbm10 at 12pt
\def\bb#1{\hbox{\mybb#1}}

\def\R {\bb{R}}


\tolerance=10000
\input phyzzx

\def\unit{\hbox to 3.3pt{\hskip1.3pt \vrule height 7pt 
width .4pt \hskip.7pt
\vrule height 7.85pt width .4pt \kern-2.4pt
\hrulefill \kern-3pt
\raise 4pt\hbox{\char'40}}}

\REF\HTa{C.M. Hull and P.K. Townsend, Nucl. Phys. {\bf B438} (1995) 109.}
\REF\PKTa{P.K. Townsend, {\it p-Brane Democracy}, hep-th/9507048, 
to appear in proceedings of the PASCOS/Hopkins workshop, March 1995.}
\REF\Wit{E. Witten, Nucl. Phys. {\bf B443} (1995) 85.}
\REF\HTb{C.M. Hull and P.K. Townsend, Nucl. Phys. {\bf B451} (1995) 525.}
\REF\Strom{A. Strominger, Nucl. Phys. {\bf B451} (1995) 96.}
\REF\Asp{P. Aspinwall, Phys. Lett. {\bf 357B} (1995) 329.}
\REF\Vafa{S. Kachru and C. Vafa, Nucl. Phys. {\bf B450} (1995) 69.}
\REF\DGHR{A. Dabholkar, G.W. Gibbons, J.A. Harvey and F. Ruiz-Ruiz,  Nucl.
Phys. {\bf B340} (1990) 33.}
\REF\CHSa{C.G. Callan, J.A. Harvey and A. Strominger, Nucl. Phys. 
{\bf B359} (1991) } 
\REF\DL{M.J. Duff and J.X. Lu, Nucl. Phys. {\bf B354} (1991) 141.}
\REF\HS{G.T. Horowitz and A. Strominger, Nucl. Phys. {\bf B360} (1991) 197.}
\REF\CHSb{C.G. Callan, J.A. Harvey and A. Strominger, Nucl. Phys. {\bf B367}
(1991) 60.}
\REF\DLb{M.J. Duff and J.X. Lu, Phys. Lett. {\bf 273B} (1991) 409.}
\REF\Ach{A. Ach\'ucarro, J.M. Evans, P.K. Townsend and D.L. Wiltshire, Phys.
Lett. {\bf 198B} (1987) 441.}
\REF\Nep{R.I. Nepomechie, Phys. Rev. {\bf D31} (1985) 1921.}
\REF\GGP{G.W. Gibbons, M.B. Green and M.J. Perry, {\it Instantons and
Seven-Branes in Type II Superstring Theory}, hep-th/9511080.}
\REF\PW{J. Polchinski and E. Witten, {\it Evidence for heterotic-Type I 
string duality}, hep-th/9510169.}
\REF\BGPT{E. Bergshoeff, M. de Roo, M.B. Green, G. Papadopoulos and P.K.
Townsend, {\it Duality of Type II 7-branes and 8-branes}, hep-th/9601150.}
\REF\GHT{G.W. Gibbons, G.T. Horowitz and P.K. Townsend, Class. Quantum Grav.
{\bf 12} (1995) 297.}
\REF\polch{J. Polchinski, Phys. Rev. Lett. {\bf 75} (1995) 4724.}
\REF\DLP{J. Dai, R.G. Leigh and J. Polchinski, Mod. Phys. Lett. {\bf A4} 
(1989) 2073.}
\REF\Leigh{R.G. Leigh, Mod. Phys. Lett. {\bf A4} (1989) 2767.}
\REF\PKTb{P.K. Townsend, Phys. Lett. {\bf 350B} (1995) 184.}
\REF\DHIS{M.J. Duff, P.S. Howe, T. Inami and K.S. Stelle, Phys. Lett.  
{\bf 191B} (1987) 70.}
\REF\DS{M.J. Duff and K.S. Stelle, Phys. Lett. {\bf 253B} (1991) 113.}
\REF\DGT{M.J. Duff, G.W. Gibbons and P.K. Townsend, Phys. Lett. {\bf 332B} (1994)
321.}
\REF\Gu{R. G\"uven, Phys. Lett. {\bf 276B} (1992) 49.}
\REF\PKTc{P.K. Townsend, Phys. Lett. {\bf 354B} (1995) 247.}
\REF\dyonm{J.M. Izquierdo, N.D. Lambert, G. Papadopoulos and P.K. Townsend,
Nucl. Phys. {\bf 460} (1996) 560.}
\REF\BST{E. Bergshoeff, E. Sezgin and P.K. Townsend, Phys. Lett. {\bf 189B}
(1987) 75; Ann. Phys. (N.Y.) {\bf 185} (1988) 330.}
\REF\HW{P. Ho{\v r}ava and E. Witten, {\it Heterotic and Type I String 
Dynamics from Eleven Dimensions}, hep-th/9510209.}
\REF\Sch{J.H. Schwarz, {\it The Power of M-theory}, hep-th/9510086.}
\REF\Schm{C. Schmidhuber, {\it D-brane Actions}, hep-th/9601003.}
\REF\Witb{E. Witten, {\it Some Comments on String Dynamics}, hep-th/9507012.}
\REF\Duff{M.J. Duff and J.X. Lu, Nucl. Phys. {\bf B416} (1994) 301.}
\REF\Hoppe{J. Hoppe, Ph.D. Thesis, MIT (1982).}
\REF\WHN{B. de Wit, J. Hoppe and H. Nicolai, Nucl. Phys. {\bf B305} [FS 23]
(1988) 545.}
\REF\Witc{E. Witten, {\it Bound states of D-strings and p-branes},
hep-th/9510135.}
\REF\BLT{E. Bergshoeff, L.A.J. London and P.K. Townsend, Class. Quantum 
Grav. {\bf 9} (1992) 2545.}
\REF\memrev{P.K. Townsend, {\it Three Lectures on Supermembranes}, in  
{\it Superstrings '88}, eds. M. Green, M. Grisaru, R.  Iengo, E. Sezgin 
and A. Strominger, (World Scientific 1989).}
\REF\GT{G.W. Gibbons and P.K. Townsend, Phys. Rev. Lett. {\bf 71} 
(1983) 3754.}
\REF\KM{D. Kaplan and J. Michelson, {\it Zero modes for the D=11 membrane 
and five-brane}, hep-th/9510053.}
\REF\BJO{E. Bergshoeff, B. Janssen and T.  Ort{\'{\i}}n, Class. Quantum Grav.
{\bf 12} (1995) 1.}
\REF\CK{C.G. Callan and I.R. Klebanov, {\it D-brane Boundary State Dynamics},
hep-th/9511173.}
\REF\MD{M. Douglas, {\it Branes within Branes}, hep-th/9512077.}
\REF\dWLN{B. de Wit, M. L\"uscher and H. Nicolai, Nucl. Phys. {\bf B320} 
(1989) 135.}
\REF\CB{C. Bachas, {\it D-brane Dynamics}, hep-th/9511043.}


\Pubnum{ \vbox{ \hbox{R/95/59} \hbox{hep-th/9512062} } }
\pubtype{}
\date{Dec. 1995, revised Jan. 1996}

\titlepage

\title {\bf D-branes from M-branes}

\author{P.K. Townsend}
\address{DAMTP, Silver St., 
\break
Cambridge CB3 9EW, U.K.}

\abstract{The 2-brane and 4-brane solutions of ten dimensional 
IIA supergravity have a dual interpretation as Dirichlet-branes, 
or `D-branes', of type IIA superstring theory and as `M-branes' 
of an $S^1$-compactified eleven dimensional supermembrane theory, 
or M-theory. This eleven-dimensional connection is 
used to determine the ten-dimensional Lorentz covariant 
worldvolume action for the Dirichlet super 2-brane, and its 
coupling to background spacetime fields. It is further used to 
show that the 2-brane can carry the
Ramond-Ramond charge of the Dirichlet 0-brane as a 
topological charge, and an interpretation of the 2-brane as 
a 0-brane condensate is suggested. Similar
results are found for the Dirichlet 4-brane via its 
interpretation as a double-dimensional reduction of the 
eleven-dimensional fivebrane. It is suggested that the latter 
be interpreted as a D-brane of an open
eleven-dimensional supermembrane.}

\endpage


\chapter{Introduction}

The importance of super p-branes for an understanding of the 
non-perturbative dynamics of type II superstring theories is 
no longer in doubt. For example, they
are relevant to U-duality of toroidally-compactified type II 
superstrings [\HTa,\PKTa], and symmetry enhancement at singular 
points in the moduli space of $K_3$ or Calabi-Yau compactified 
type II superstrings [\Wit,\HTb,\Strom,\Asp] as required by the 
type II/heterotic string-string duality [\HTa,\Wit,\Vafa]. Type II
p-branes were first found as solutions of the effective D=10 
supergravity theory [\DGHR,\CHSa,\DL,\HS]. Because their 
worldvolume actions involve worldvolume gauge
fields [\CHSb,\DLb], in addition to the scalars and spinors 
expected on the basis of spontaneously broken translation
invariance and supersymmetry, they were not anticipated in the 
original classification of super p-branes [\Ach]. For the 
same reason, the fully D=10 Lorentz covariant action for these 
type II super p-branes is not yet known. One purpose of this paper
is to report progress on this front.

The type II p-branes are conveniently divided into those of
Neveu/Schwarz-Neveu/Schwarz (NS-NS) type and those of Ramond-Ramond 
(RR) type according to the string theory origin of the (p+1)-form gauge potential for which they are a source. The supergravity super p-branes 
found in the NS-NS sector comprise a string and a fivebrane. The string 
has a naked timelike singularity and can be identified as the effective 
field theory realization of the fundamental string\foot{Note that the 
existence of this solution is necessary for the
consistency of any string theory with massless spin 2 excitations 
since a macroscopic string will then have long range fields which must 
solve the source free equations of the effective field theory.}. The 
fivebrane solution is non-singular and has a 5-volume tension $\sim \lambda^{-2}$ expected of a soliton, where $\lambda$ is the string 
coupling constant. Since a 5-brane is the magnetic dual of a string 
in D=10 [\Nep], this solution is an analogue of the BPS magnetic 
monopole of D=4 super Yang-Mills (YM) theory.

In the RR sector the ten-dimensional (D=10) IIA supergravity has p-brane solutions for p=0,2,4,6, while the IIB theory has RR p-branes solutions 
for p=1,3,5 (see [\PKTa] for a recent review)\foot{There is also a IIB 
7-brane [\GGP] and a IIA 8-brane [\PW] (see also [\BGPT]), but these 
do not come in electric/magnetic pairs and have rather different physical implications; for example, they do not contribute to the spectrum of 
particles in any $D\ge 4$ compactification. Partly for this reason, only 
the $p\le6$ cases will be discussed here.}. With the exception of the 
3-brane, which is self-dual, these p-branes come in
$(p,\tilde p)$ electric/magnetic pairs with $\tilde p=6-p$ . The RR 
p-brane solutions all have a p-volume tension $\sim \lambda^{-1}$ 
[\Wit] so, although non-perturbative, they are not typically solitonic. Moreover, they are all singular, with the exception of the 3-brane, and even this exceptional case is not typical of solitons because the solution has an event horizon [\GHT]. Thus, the RR p-branes are intermediate between the fundamental string and the solitonic fivebrane. It now appears [\polch] that they have their place in string theory as Dirichlet-branes, or D-branes [\DLP,\Leigh].

It was shown in [\PKTb] how all the p-brane solutions of D=10 
IIA supergravity (with $p\le 6$) have an interpretation in D=11, 
extending previous results along these lines for the string and fourbrane [\DHIS,\DS,\DGT]. In particular, the 0-branes were identified with the Kaluza-Klein (KK) states of D=11 supergravity and their 6-brane duals 
were shown to be D=11 analogues of the KK monopoles. The remaining 
p-brane solutions have their D=11 origin in either the membrane [\DS]
or the fivebrane [\Gu] solutions of D=11 supergravity. It was 
subsequently shown that D=11 supergravity is the effective field 
theory of the type IIA superstring at strong coupling [\Wit] and 
then that various dualities in $D<10$ can be understood in terms of the electric/magnetic duality in D=11 of the membrane and fivebrane 
[\PKTc,\dyonm]. These results suggest the existence of a consistent 
quantum theory underlying D=11 supergravity. This may be a supermembrane 
theory as originally suggested [\BST], or it may be some
other theory that incorporates it in some way. Whatever it is, it 
now goes by the name `M-theory' [\HW,\Sch]. 

The point of the above summary is to show that the RR p-brane 
solutions of D=10 IIA supergravity theory currently have two 
quite different interpretations. On the one hand they are interpretable 
as D-branes of type IIA string theory. On the other hand they are 
interpretable as solutions of $S^1$ compactified D=11 supergravity. 
In the $p=2$ and $p=4$ cases these D=11 solutions are also 
p-branes; since they are presumably also solutions of the underlying  
D=11 M-theory we shall call them `M-branes'. We shall first exploit the 
interpretation of the $p=2$ super D-brane as a dimensionally reduced 
D=11 supermembrane to deduce its D=10 Lorentz covariant worldvolume 
action. The bosonic action has been found previously [\Leigh] by requiring one-loop conformal invariance of the open string with the string 
worldsheet boundary on the D-brane\foot{The action of [\Leigh] is 
not obviously equivalent to the bosonic sector of the one found here 
and the omission of a discussion of this point was a defect of an earlier version of this paper; fortunately, the equivalence has
since been established by Schmidhuber [\Schm].}. One feature of the 
derivation via D=11 is that the coupling to background fields can 
also be found this way, and the resulting action has a straightforward generalization to general $p$. The coupling to the dilaton is such that the p-volume tension is $\sim\lambda^{-1}$, as expected for a D-brane [\DLP]. 
The M-brane interpretation of the Dirichlet 4-brane is as the
{\it double}-dimensional reduction of the eleven-dimensional fivebrane. 
We propose a bosonic action for the latter including a coupling to 
the bosonic fields of eleven-dimensional supergravity, and exploit 
it to deduce the coupling to background supergravity fields, 
including the dilaton, of the Dirichlet 4-brane. The result agrees 
with that deduced by generalization of the $p=2$ case.

One intriguing feature of these results is that they suggest an 
interpretation of
the eleven-dimensional fivebrane as a Dirichlet-brane of an 
open D=11 supermembrane, and we further suggest that the 
string-boundary dynamics is
controlled by the conjectured [\Witb], and intrinsically 
non-perturbative, six-dimensional self-dual string theory 
(which is possibly related to the 
self-dual string soliton [\Duff], although this solution involves 
six-dimensional gravitational fields which are not, according to
current wisdom, among the fivebrane's worldvolume fields).

Finally, we show that a spherical D=10 2-brane can carry the 
same RR charge that is carried by the Dirichlet 0-branes; this 
charge is essentially the magnetic
charge associated with the worldvolume vector potential. 
This suggests that the 0-branes can be viewed as collapsed 
2-branes. We point out that this is
consistent with the $U(\infty)$ Supersymmetric Gauge 
Quantum Mechanics interpretation of the supermembrane worldvolume 
action [\Hoppe,\WHN], which further suggests an interpretation 
of the supermembrane as a condensate of
0-branes. Viewed from the D=11 perspective these results can 
be taken as further evidence that D=11 supergravity is the 
effective field theory of a supermembrane theory.


\chapter{The D=10 2-brane as a D=11 M-brane}

Consider first the D=10 2-brane. From its D-brane description 
we know that the worldvolume action is based on the D=10 Maxwell 
supermultiplet dimensionally reduced to three dimensions [\Witc], 
i.e. the worldvolume field content is
$$
\{ X^a\  (a=1,\dots,7)\; ,\; A_i\ (i=0,1,2)\;  ;\chi^I\ (I=1,\dots,8)\}
\eqn\aone
$$
where the $\chi^I$ are eight $Sl(2;\R)$ spinors and $A_i$ is a 
worldvolume vector
potential\foot{Throughout this paper we shall use the letter 
$A$ to denote worldvolume gauge fields, of whatever rank, and 
$B$ to denote spacetime gauge fields, of whatever rank.}. As for 
every other value of $p$, only the bosonic part of the 10-dimensional 
Lorentz covariant action constructed from these fields
is currently known [\Leigh]. However, the alternative interpretation 
of the 2-brane as an M-brane allows us to find the complete action. 
In this interpretation, the IIA 2-brane is the direct (as against double) dimensional reduction of the D=11 supermembrane. The worldvolume fields 
of the dimensionally reduced D=10 supermembrane are, before 
gauge-fixing, $\{X^m\ (m=0,1,\dots,9);\,\varphi;\,\theta\}$, where 
$\theta$ is a 32-component Majorana spinor of the D=10 Lorentz 
group and $X^m$ is a 10-vector. After gauge
fixing the physical fields are
$$
\{ X^a\  (a=1,\dots,7)\; ,\; \varphi\, ; \chi^I\ (I=1,\dots,8)\}\ .
\eqn\bone
$$
The difference between \aone\ and \bone\ is simply that 
the scalar $\varphi$ of \bone\ is replaced in \aone\ by its 
3-dimensional dual, the gauge vector $A$. By
performing this duality transformation in the action 
{\it prior} to gauge fixing we can determine the fully 
D=10 Lorentz covariant Dirichlet supermembrane action.

The first step of this procedure is to isolate the 
dependence of the D=11 supermembrane action on $X^{11}$, 
which is here called $\varphi$. 
We shall first consider the case for which the D=11 
spacetime is the product of $S^1$ with D=10 Minkowski 
spacetime, returning subsequently to consider the
interaction with background fields. It is convenient 
to use the Howe-Tucker (HT) formulation of the action for 
which there is an auxiliary worldvolume
metric $\gamma_{ij}$. It is also convenient to introduce 
the spacetime supersymmetric differentials
$$
\Pi^m =dX^m - i\bar\theta\Gamma^md\theta\ .
\eqn\susydiff
$$
The action, given in [\BST], is
$$
\eqalign{
S= -{1\over2}\int\! &d^3\xi\,
\sqrt{-\gamma}\big[\gamma^{ij}\Pi^m_i\Pi^n_j\eta_{mn} +
\gamma^{ij}(\partial_i\varphi -i\bar\theta\Gamma_{11}\partial_i\theta)
(\partial_j\varphi -i\bar\theta\Gamma_{11}\partial_j\theta) -1\big]\cr
& -{1\over6} \int\! d^3\xi\; \varepsilon^{ijk}[b_{ijk}
+3 b_{ij}\partial_k\varphi]\ , }
\eqn\cone
$$
where $\eta$ is the D=10 Minkowski metric, and
$$
\eqalign{
\varepsilon^{ijk}b_{ijk} = 
3\varepsilon^{ijk}\Big\{ &i\bar\theta \Gamma_{mn}\partial_i\theta \big[
\Pi_i^m\Pi_j^n\eta_{mn} + i\Pi_i^m(\bar\theta\Gamma^n \partial_j\theta)
-{1\over3}(\bar\theta\Gamma^m\partial_i\theta)
(\bar\theta\Gamma^n\partial_j\theta)\big]\cr
&+ (\bar\theta\Gamma_{11}\Gamma_m\partial_i\theta)
(\bar\theta\Gamma_{11}\partial_j\theta)
(\partial_kX^m-{2i\over3}\bar\theta\Gamma^m\partial_k\theta)\Big\} \ ,}
\eqn\bijk
$$
while
$$
\varepsilon^{ijk}b_{ij}=
-2\varepsilon^{ijk}\; i\bar\theta\Gamma_m\Gamma_{11}\partial_i\theta
(\partial_jX^m -{i\over2}\bar\theta\Gamma^m\partial_j\theta) \ .
\eqn\bij
$$

The second step, the replacement of the worldvolume scalar $\varphi$ 
by its dual vector field, can be achieved by promoting $d\varphi$ 
to the status of an independent worldvolume one-form $L$ while adding 
a Lagrange multiplier term $AdL$ to impose the constraint $dL=0$. 
Eliminating $L$ by its algebraic equation
of motion yields the dual action in terms of the fields $X^m$ and the
worldvolume field strength two-form $F=dA$. This action is
$$
\eqalign{
S = -{1\over2}\int \! d^3\xi\; &\sqrt{-\gamma}\Big[
\gamma^{ij}\;\Pi^m_i\Pi^n_j\;
\eta_{mn} + {1\over2}\gamma^{ik}\gamma^{jl} 
\hat F_{ij}\hat F_{kl} -1\Big] \cr
&-{1\over6}\int\! d^3\xi\;\varepsilon^{ijk} \big[b_{ijk}
-3i(\bar\theta\Gamma_{11}\partial_i\theta)\hat F_{jk}\big]\ .}
\eqn\done
$$
where
$$
\hat F_{ij} = F_{ij} -b_{ij}\ .
\eqn\donea
$$
Thus \done\ is the fully D=10 Lorentz covariant worldvolume 
action for the D=10 IIA Dirichlet supermembrane. The bosonic 
action, obtained by setting the fermions to zero in \done, is 
equivalent to the Born-Infeld-type action found by Leigh [\Leigh]. 
The equivalence follows from the recent observation of
Schmidhuber [\Schm] that dualizing the vector to a scalar in the 
action of Leigh yields the action of a D=11 membrane, which was 
precisely the (bosonic) starting
point of the construction presented here\foot{The equivalence
with Born-Infeld for p=1, i.e. the D-string, was shown in [\BLT]}.
It is interesting to note that a
sigma-model one-loop calculation in the string theory is 
reproduced by the classical supermembrane. 

It can now be seen why it was advantageous to start from the
HT form of the action; whereas the auxiliary metric is simply
eliminated from \cone, leading to the standard 
Dirac-Nambu-Goto (DNG) form of the action, its elimination 
from \done\ is far from straightforward, although
possible in principle. The point is that the $\gamma_{ij}$ 
equation is now the very non-linear, although still algebraic, 
equation
$$
\gamma_{ij} = \Big(1+ {1\over2} \gamma^{kp}\gamma^{lq}
\hat F_{kl} \hat F_{pq}\Big)^{-1} \Big( g_{ij} + 
\gamma^{kl}\hat F_{ik} \hat
F_{lj}\Big)
\eqn\doneb
$$
where $g_{ij}=\Pi_i^m\Pi_j^n\;\eta_{mn}$. This equation 
can be solved as a series in $\hat F$ of the form
$$
\gamma_{ij} = g_{ij}\big[ 1-{1\over2} g^{kp} g^{lq} 
\hat F_{kl}\hat F_{pq}\big] 
+ g^{kl} \hat F_{ik} \hat F_{jl} + {\cal O}(\hat F^4)\ ,
\eqn\eone
$$
and the approximation $\gamma_{ij}=g_{ij}$ yields the 
quadratic part of the action in $\hat F$.

Invariance of the action \done\ under supersymmetry requires $\hat
F$ to be invariant. To see how this comes about, we observe that 
the two-form $b$ in $\hat F$ is precisely the one that defines the 
WZ term in the Green-Schwarz superstring action; it has the property 
that the three-form $h=db$ is superinvariant, which implies that $\delta_\epsilon b=da$ for some one-form $a(\epsilon)$, where 
$\epsilon$ is the (constant) supersymmetry parameter. The 
modified two-form field strength $\hat F$ is therefore superinvariant 
if we choose $\delta_\epsilon A = a$. The $\kappa$-transformation of 
$A$ is similarly determined by requiring $\kappa$-gauge invariance 
of the action, but it can also be deduced directly from
those of the D=11 supermembrane given in [\BST]. The result is 
most simply expressed in terms of the variations of the 
supersymmetric forms $\Pi^m$ and $\hat F$, which 
are\foot{As explained in detail in [\memrev], it
is not necessary to specify the transformation of the 
metric $\gamma_{ij}$ if use is made of the `1.5 order' formalism.}
$$
\eqalign{
\delta_\kappa \Pi^m &= -2i(\delta_\kappa\bar\theta) 
\Gamma^m d\theta \cr
\delta_\kappa \hat F &= i (\delta_\kappa \bar\theta)
\Gamma_m\Gamma_{11} d\theta \wedge \Pi^m \cr
\delta_\kappa \theta &= \big( 1+\Gamma)\kappa\ ,}
\eqn\kaptran
$$
where
$$
\Gamma = {1\over 6\sqrt{-\gamma}}\varepsilon^{ijk} 
\Pi_i^m\Pi_j^n\Pi_k^p
\Gamma_{mnp} -{1\over2} 
\gamma^{ik}\gamma^{jl}F_{kl}\Pi_i^m\Pi_j^n\Gamma_{mn}
\Gamma_{11}
\eqn\akap
$$
and $\kappa(\xi)$ is the D=10 Majorana spinor parameter.

The coupling of the action \done\ to background fields 
can also be deduced from its D=11 origin. We shall consider 
here only the bosonic membrane coupled to bosonic background 
fields. Consider first the  NS-NS fields. In the D=10 membrane 
action obtained by direct dimensional reduction from D=11, 
the NS-NS two-form potential $B$ couples to the topological
current $\varepsilon^{ijk}\partial_k\varphi$. In the dual 
action this coupling corresponds to the replacement of $F$ 
by $F-B$. The coupling to the D=10 spacetime metric is obvious 
so this leaves the dilaton; to determine its 
coupling we recall (see e.g. [\DGT]) that the D=11 metric is
$$
ds^2_{11} = e^{-{2\over3}\phi} ds^2 + 
e^{{4\over3}\phi}d\varphi^2\ ,
\eqn\fone
$$
where $ds^2$ is the string-frame D=10 metric and 
$\phi$ is the dilaton. A repetition of the steps described 
above, but now for the purely bosonic theory and carrying 
along the dependence on the NS-NS-spacetime fields, leads 
(after a redefinition of the auxiliary metric to
the action
$$
S= -{1\over2}\int \! d^3\xi\; e^{-\phi}\;
\sqrt{-\gamma}\Big[\gamma^{ij}g_{ij} +
{1\over2}\gamma^{ik}\gamma^{jl} 
(F_{ij}-B_{ij})( F_{kl}-B_{kl}) -1\Big] \ ,
\eqn\gonea
$$
where now $g_{ij}=\partial_iX^m\partial_jX^ng_{mn}$. 
The appearance of $F$ through the modified field strength $F-B$ 
could also have been deduced simply by the replacement of the 
flat superspace two-form potential $b$ in $\hat F$ by its curved 
superspace counterpart, since setting the fermions to
zero then yields precisely $F-B$. As for the RR fields, the
coupling to the 3-form potential is of course the standard 
Lorentz coupling while the coupling to the 1-form potential 
has interesting implications which
will be discussed at the conclusion of this article.

The above result, and the known form of the bosonic 
p-brane action in the absence of worldvolume gauge fields, 
suggests that the corresponding
bosonic part of the worldvolume action of the Dirichlet 
super p-brane is
$$
S= -{1\over2}\int \! d^{(p+1)}\xi\; e^{-\phi}\;
\sqrt{-\gamma}\Big[\gamma^{ij}g_{ij} +
{1\over2}\gamma^{ik}\gamma^{jl} 
(F_{ij}-B_{ij})(F_{kl}-B_{kl}) -(p-1)\Big] \ .
\eqn\ione
$$
Since the vacuum expectation value of $e^\phi$ is the string 
coupling constant $\lambda$, it follows from this result that the 
$p$-brane tension is $\sim \lambda^{-1}$, as expected for 
D-branes. Of course, the steps leading to this result were 
particular to $p=2$ but we shall shortly arrive at the same 
result for $p=4$ via a different route. Although the action 
\ione\ is only guaranteed to be correct to quadratic order in 
$F$ for $p\ne2$, this will prove sufficient
for present purposes.


\chapter{The D=11 5-brane as a supermembrane D-brane}

Consider now the Dirichlet 4-brane. In this case its 
M-brane interpretation is as a double-dimensional reduction 
of the D=11 5-brane. The (partially)
gauge-fixed field content of the latter consists 
[\GT,\KM] of the fields of the
N=4 six-dimensional antisymmetric tensor multiplet, i.e.
$$
\{ X^a\ (a=1,\dots,5)\; ,\; A^+_{ij}\ (i,j=0,1,\dots,5)\; ; \chi^I (I=1,\dots,4)\}
\eqn\athree
$$
where $\chi^I$ are chiral symplectic-Majorana spinors 
in the ${\bf 4}$ of $USp(4)\cong Spin(5)$, and $A^+$ is 
the two-form potential for a self-dual
3-form field strength $F=dA^+$. Because of the self-duality 
of $F$ we cannot expect to find a worldvolume action (at least, 
not one quadratic in $F$). We  might try to find an action 
that leads to all equations {\it except} the
self-duality constraint which we can then just impose by 
hand, as advocated elsewhere in another context [\BJO]. 
We shall adopt this strategy here, but it
is important to appreciate an inherent difficulty 
in its present application.
The problem is that the self-duality condition involves a
metric and it is not clear which metric should be used, 
e.g. the induced metric or the auxiliary
metric; the possibilities differ by higher order terms 
in $F$. Because of this ambiguity we should consider the 
action as determining only the lowest order,
quadratic, terms in $F$. With this proviso, an obvious 
conjecture for the D=11 5-brane action is
$$
S=-{1\over2}\int\! d^6\xi\; 
\sqrt{-\gamma}\Big[ \gamma^{ij}\partial_i
X^M\partial_j X^N\eta_{MN} + 
{1\over2}\gamma^{il}\gamma^{jm}\gamma^{kn}
F_{ijk}F_{lmn} -4\Big]\ ,
\eqn\bthree
$$
where the fields $X^M$,$\; M=(0,1,\dots,10)$, are 
maps from the worldvolume to the D=11 Minkowski spacetime. 
This action has an obvious coupling to the bosonic fields
$(g_{MN}, B_{MNP})$ of D=11 supergravity\foot{although 
consistency with the self-duality condition is now problematic.}. 
The coupled action is
$$
S=-{1\over2}\int\! d^6\xi\; \sqrt{-\gamma}
\Big[\gamma^{ij}g_{ij}^{(11)} +
{1\over2} \gamma^{il}\gamma^{jm}\gamma^{kn} 
\big(F_{ijk}-B_{ijk}\big)
\big(F_{lmn}-B_{lmn}\big) -4\Big]
\eqn\cthree
$$
where $g_{ij}^{(11)}$ is the pullback of the 11-metric $g_{MN}$ 
and $B_{ijk}$ is the pullback of the 3-form potential $B_{MNP}$. 
Up to quartic terms in $F_{ij}$, and setting to zero the RR 
spacetime fields, the double dimensional reduction of \cthree\ to 
D=10 reproduces the action \ione\ with p=4, as required
for the M-brane interpretation of the Dirichlet 4-brane. In 
particular, the dilaton dependence is exactly as given in \ione.

The worldvolume vector of the D=10 Dirichlet p-branes allows 
not only a coupling to the 2-form potential of string theory 
but also to the endpoints of an open
string via a boundary action [\DLP,\Leigh,\CK]. Let 
$X^m(\sigma,\tau)$ be the locus in spacetime of the string's 
worldsheet, with boundary at $\tau=0$. If
this boundary lies in the worldvolume of a $p$-brane, then
$$
X^m(\sigma,\tau)\Big|_{\tau=0} = X^m\big(\xi(\sigma)\big)\ ,
\eqn\extraone
$$
where $X^m(\xi)$ is the locus in spacetime of the $p$-brane's 
worldvolume. It is also convenient to introduce the conjugate 
momenta to the worldsheet scalar fields at the worldsheet 
boundary, $\pi_m$, defined by 
$$
\pi_m(\sigma) = \sqrt {-g}\; g_{mn}\big(X(\sigma,\tau)\big)\;
{dX^n(\sigma,\tau)\over d\tau}\Big|_{\tau=0}\ .
\eqn\extratwo
$$
The D=10 Lorentz covariant boundary action can then be 
written as 
$$
S_b(string) = \oint \! d\sigma\Big[A_i\big(\xi(\sigma)\big) 
{d\xi^i(\sigma)\over
d\sigma}\  + \ X^m\big(\xi(\sigma)\big) \pi_m(\sigma) \Big]\ .
\eqn\dthree
$$

Similarly, the worldvolume antisymmetric tensor $A^+$ of the 
D=11 5-brane allows not only a coupling to the 3-form potential 
of D=11 supergravity but also to the boundary of an open membrane. 
Let $X^M(\sigma,\rho,\tau)$ be the locus in the D=11 spacetime 
of the membrane's worldvolume, with boundary at $\tau=0$. If
this boundary lies in the worldvolume of a fivebrane, with 
coordinates $\xi^i$, then
$$
X^M(\sigma,\rho,\tau)\Big|_{\tau=0} = 
X^M\big(\xi(\sigma,\rho)\big)\ ,
\eqn\extrathree
$$
where $X^M (\xi)$ is the locus in spacetime of the 
fivebrane's worldvolume. Defining, as before, the conjugate 
momenta $\pi_M$ to the membrane scalar fields
at the membrane's boundary, we can write down the 
following natural generalization of \dthree:
$$
S_b(membrane) = \oint \! d\sigma d\rho \Big[A^+_{ij}(\xi)
{d\xi^i\over d\sigma}{d\xi^j\over d\rho}\  + 
\ X^m(\xi) \pi_M
\Big]\ .
\eqn\ethree
$$
Moreover, the double-dimensional reduction of this membrane 
boundary action reproduces the string boundary action \dthree. 
This suggests that we interpret the D=11 5-brane as a 
Dirichlet-brane of an underlying open
supermembrane. It seems possible that the dynamics of the 
membrane boundary in the fivebrane's worldvolume might be 
describable by a six-dimensional superstring theory, which 
one would expect to have N=2 (i.e. minimal) six-dimensional
supersymmetry (e.g. on the grounds that it is a `brane within 
a brane' [\MD]). However, since the 3-form field strength to 
which this boundary string couples is self-dual, this superstring 
theory would be, like the supermembrane itself, intrinsically 
non-perturbative. The existence of such a new superstring theory 
was conjectured previously [\Witb] in a rather different context.


\chapter{0-branes from 2-branes and 2-branes from 0-branes.}

One of the properties expected of the D=11 supermembrane 
theory or M-theory is that it have D=11 supergravity as its 
effective field theory. Various arguments for and against this 
have been given previously ([\PKTb] contains a recent
brief review). A further argument in favour of this idea is 
suggested by the recent results of Witten concerning the 
effective action of $n$ coincident Dirichlet $p$-branes [\Witc]. 
He has shown that the (partially gauge-fixed) effective action 
in this case is the reduction from D=10 to $(p+1)$ dimensions 
of the $U(n)$ D=10 super Yang-Mills (YM) theory. Consider the 
0-brane case for which the super YM theory is one-dimensional 
i.e. a model of supersymmetric gauge quantum mechanics (SGQM). 
If the 0-branes condense at some point then the
effective action will be the $n\rightarrow \infty$ limit of a 
$U(n)$ SGQM. But this is just another description of the 
supermembrane! It is amusing to note that the continuity of the 
spectrum of the quantum supermembrane [\dWLN], in 
the zero-width approximation appropriate to its D-brane 
description, might now be understood as a consequence of the 
zero-force condition between an infinite number of 
constituent 0-branes. However, it is known that quantum 
string effects cause the D-brane to aquire a finite size 
core [\CB], consistent with its M-brane interpretation as a 
solution with an event horizon [\DGT], and it was argued in 
[\PKTb] that this fact should cause the spectrum to be discrete.

Actually, the supermembrane was usually stated as 
being equivalent to an $SU(\infty)$ SGQM model [\Hoppe,\WHN], 
but the additional $U(1)$ is needed to describe the dynamics 
of the centre of mass motion. Note that a $U(1)$ SGQM
is precisely the action for a Dirichlet 0-brane. This suggests 
that there might exist some classical closed membrane configuration 
for which the ground state, on quantization, could be identified 
with the 0-brane. For this to be possible
it would be necessary for the closed membrane to carry the
RR charge associated with the 0-branes. We now explain how 
this can occur.

From the D=11 point of view the RR 0-brane charge is just the 
KK charge, i.e. the electric charge that couples to the
KK vector field, which we shall here call $B_m$. The coupling 
of $B_m$ to the D=10 membrane can be found by dimensional 
reduction from D=11. To leading order this coupling has the 
standard Noether form $B_m{\cal J}^m$, where
$$
{\cal J}^m(x) = \int\! d^3\xi\;\sqrt{-\gamma}\gamma^{ij}\partial_i
X^m\partial_j\varphi\, \delta^{10}\big(x-X(\xi)\big)\ .
\eqn\afive
$$
is the KK current density. After dualization of the scalar 
field this becomes
$$
{\cal J}^m(x) = \int\! d^3\xi\;\varepsilon^{ijk}\partial_i
X^m F_{jk}\, \delta^{10}\big(x-X(\xi)\big)\ .
\eqn\bfive
$$
The total KK charge is $Q\equiv \int d^9x {\cal J}^0$. 
Choosing the $X^0=\xi^0$
gauge one readily sees that
$$
Q = \oint F\ ,
\eqn\cfive
$$
i.e. the integral of the worldvolume 2-form field 
strength $F$ over the
closed membrane. 

Thus, a closed membrane can carry the 0-brane RR charge 
as a type of magnetic charge associated with its 
worldvolume vector field, and its centre of 
mass motion is described by the 0-brane $U(1)$ SQGM. 
This can be interpreted as further evidence that the 
0-brane is included in the (non-perturbative) 
supermembrane spectrum. However, from the D=11 
point of view the 0-brane is just a massless quantum of 
D=11 supergravity and supersymmetry implies the existence 
of all massless quanta given any one of them. Thus, we have 
found a new argument that the spectrum of the D=11 
supermembrane (or, perhaps, M-theory) should include the 
massless states of D=11 supergravity.

\vskip 0.5cm
\noindent 
{\bf Acknowledgements}: Helpful conversations with 
M.B. Green, J. Hoppe, R.G. Leigh, G. Papadopoulos and 
A. Tseytlin are gratefully acknowledged.

\refout

\end